\documentclass[pra,showpacs,twocolumn,superscriptaddress,showkeys]{revtex4}

\usepackage{psfrag}
\usepackage{graphicx}

\begin{document}

\title{Entanglement and Quantum Phases in the Anisotropic Ferromagnetic Heisenberg 
Chain in the Presence of Domain Walls}
\author{F.C. \surname{Alcaraz}}
\email{alcaraz@if.sc.usp.br}
\affiliation{Instituto de F\'{\i}sica de S\~ao Carlos, Universidade de S\~ao Paulo, \\
Caixa Postal 369, 13560-590, S\~ao Carlos, S\~ao Paulo, Brazil. \vspace{0.1cm}}
\author{A. \surname{Saguia}}
\email{saguia@spinon.uwaterloo.ca}
\affiliation{Department of Physics, University of Waterloo,   \\
200 University Avenue W., Waterloo, Ontario, N2L 3G1, Canada.\vspace{0.1cm}}
\author{M.S. \surname{Sarandy}}
\email{msarandy@chem.utoronto.ca}
\affiliation{Chemical Physics Theory Group, Department of Chemistry, University of Toronto, \\
80 St. George Street, Toronto, Ontario, M5S 3H6, Canada.\vspace{0.1cm}}
\date{\today }

\begin{abstract}
We discuss entanglement in the spin-1/2 anisotropic ferromagnetic Heisenberg chain in the 
presence of a boundary magnetic field generating domain walls. By increasing the magnetic field,
the model undergoes a first-order quantum phase transition from a ferromagnetic to a kink-type phase, 
which is associated to a jump in the content of entanglement available in the system.
Above the critical point, pairwise entanglement is shown to be non-vanishing and independent of
the boundary magnetic field for large chains. Based on this result, we
provide an analytical expression for the entanglement between arbitrary spins. Moreover
the effects of the quantum domains on the gapless region and for antiferromagnetic
anisotropy are numerically analysed. Finally multiparticle entanglement properties are 
considered, from which we establish a characterization of the critical anisotropy separating the gapless regime 
from the kink-type phase.
\end{abstract}

\pacs{03.65.Ud, 03.67.Mn, 75.10.Jm}
\keywords{Entanglement, Quantum spin systems, Quantum critical phenomena}

\maketitle

\section{Introduction}

In recent years, a great deal of effort has been devoted to the understanding of
entanglement in spin systems. In fact, a number of schemes for
quantum information processing based on spin interactions have been
proposed (e.g.,~\cite{Kane:98,DiVincenzo:99,Loss:99,DiVincenzo:00,Mizel:04}), 
motivating the theoretical study of the amount 
of entanglement present in spin models~\cite{Wootters:01,Wootters:02,ved1,ved2,wxy,Wang:01,
Fu:01,Fu:04,Sch:03,Glaser:03}. 
Furthermore, the interest in the content of entanglement available in statistical systems  
has also been increased due to the observed relationship between entanglement and the theory 
of quantum critical phenomena~\cite{amico,nielsen,Vidal:03,Latorre:04,Li:03,Buzek:03,cirac,ibose,Wu:04}. 
Indeed, it has been noted that quantum phase transitions present in condensed matter systems
can be described from the point of view of entanglement. As examples, the ground state 
pairwise entanglement of XY and XXZ chains in a transverse magnetic field has been worked 
out~\cite{amico,nielsen,Vidal:03,Latorre:04,Li:03,Buzek:03}, 
from which scaling behavior near the transition point is observed. 
For the transverse field Ising model, the quantum phase transition has also been characterized 
through the concept of localizable entanglement~\cite{cirac}, which is the maximal amount of entanglement 
that can be localized for two particles in the system by doing local measurements on the other particles. 
The typical scale for which the localizable entanglement decays defines the entanglement length, which 
diverges in the Ising quantum critical point~\cite{cirac}. Moreover, in the case of first-order 
quantum phase transitions, changes in the amount of pairwise entanglement at quantum critical points 
have been shown to be associated to macroscopic jumps in the magnetization for some frustrated 
spin models~\cite{ibose}. In a general framework, the association between quantum phase transitions and 
bipartite entanglement has recently been analysed in Ref.~\cite{Wu:04}. 
Hence, in addition to its intrinsic relevance for quantum information 
applications, entanglement also plays an interesting role in the context of statistical mechanics.

Keeping both motivations in mind, the aim of this work is to analyse 
the effect of a topologically non-trivial structure on the
entanglement of a spin system. More specifically, we shall discuss the spin-1/2 anisotropic
ferromagnetic Heisenberg chain in the presence of domain walls. A related discussion about  
entanglement in spin models with open boundary conditions can be found in Ref.~\cite{Wang:04}. 
The quantum domains can be generated by an antiparallel magnetic field on the boundary spins of the chain.
For a critical value of the magnetic field, the system undergoes a
first-order quantum phase transition~\cite{alcaraz}, which arises from the interplay among the
exchange coupling $J$ in the XY direction, the anisotropy $\Delta$ in the $z$-axis ($\Delta\ge J$),
and the magnetic field $h$ on the boundary. The quantum critical point separates a ferromagnetic
phase from a kink-type phase, in which we show that entanglement is
present. It is interesting to observe that entanglement is known to be usually absent for ferromagnetic
Heisenberg interactions regardless a magnetic field is present in a particular direction or the
effect of temperature is taken into account~\cite{ved1}. Even in the anisotropic chain, namely
the XXZ model, entanglement is vanishing in the region $\Delta>J$~\cite{Wang:01,Fu:01}. Actually it has 
recently been shown in Ref.~\cite{Asoudeh:04} that entanglement can be nonvanishing for finite 
ferromagnetic Heisenberg chains at low temperatures when a magnetic field is applied. 
However this entanglement is seen to disappear in the thermodynamical limit. The domain walls can thus 
be seen, due to the induction of the phase transition, as a general mechanism to promote the existence 
of entanglement in Heisenberg models with ferromagnetic couplings. Moreover, we will show that an analysis
for large chains, in the kink-type phase, suggests that entanglement is independent of the boundary
magnetic field. This result will allow for a general expression, 
in terms of the length of the chain, for the entanglement between arbitrary spins in the ground state, 
which is exactly known at the critical point. We shall also investigate the region $\Delta<J$, 
corresponding to a gapless regime for $-J\le\Delta<J$ and a 
gapful phase for $\Delta<-J$. For these ranges of anisotropy, the distribution of nearest neighbor 
entanglement oscillates along the chain, exhibiting a clearly different behavior from the 
kink-type phase, where a monotonic increase is observed from the boundaries towards the center. 
Interestingly, a characterization of the critical anisotropy $\Delta=J$ can also be achieved 
from multiparticle entanglement properties of the model. Indeed, we will show that 
$\Delta=J$ corresponds to a minimum of the derivative of the global measure 
of entanglement introduced by Meyer and Wallach in Ref.~\cite{mw}.    

The paper is organized as follows. In the section II, we describe the quantum
Heisenberg model for a ferromagnetic domain wall, numerically discussing, for chains up
to $24$ sites, the general properties of the entanglement available in the system and its relation with the 
quantum phase transition. In Section III, we perform an analytical computation of entanglement 
for general pairs of spins starting from the exact ground state wave functions at the critical 
point. Section IV is devoted to the numerical discussion of the case $\Delta<J$. 
In Section V we analyse the entanglement properties of the model from the point of 
view of a multipartite measure. Finally, in section V, we summarize our results, 
presenting our conclusions.

\section{Anisotropic Heisenberg model for ferromagnetic domain walls}

\label{cc}

The Hamiltonian for the one-dimensional Heisenberg model in the presence of a 
magnetic field generating domain walls reads~\cite{sch,alcaraz}
\begin{eqnarray}
H &=&-J\sum_{i=1}^{L-1}(S_{i}^{x}S_{i+1}^{x}+S_{i}^{y}S_{i+1}^{y})-\Delta
\sum_{i=1}^{L-1}S_{i}^{z}S_{i+1}^{z}\hspace{0.5cm}  \nonumber \\
&&-h(S_{1}^{z}-S_{L}^{z}),\hspace{4.5cm}  \label{hdw}
\end{eqnarray}%
where the coupling $J>0$ and the anisotropy $\Delta \geq J$ are exchange parameters, and the 
effective magnetic field $h>0$ represents the interactions of the spins with the boundary
surfaces. The spin-1/2 operators $S^{\alpha}_i$, $\alpha=x,y,z$, act on the site $i$ ($i=1,...,L$) and are 
given by $S^{\alpha}=\sigma^\alpha /2$, with $\sigma^\alpha$ denoting Pauli matrices. In 
order to focus the discussion on the values of the anisotropy and the magnetic field, let us assume,
without loss of generality, $J=1$. As shown in Ref.~\cite{alcaraz}, the model presents a critical
field
\begin{equation}
h_{c}=\frac{1}{2}\sqrt{\Delta ^{2}-1}  \label{hcrit}
\end{equation}%
that separates two quantum phases for a chain of arbitrary length: a ferromagnetic ground state 
($h<h_{c}$) and kink-type ground state ($h>h_{c}$). Remarkably, Eq.~($\ref{hcrit}$) provides
the exact location of the phase transition for chains of any size, the critical 
field remaining fixed as  the number of sites is changed. 
Moreover, the ground state is exactly known at the quantum critical point, with 
degenerate wave functions given by~\cite{alcaraz}
\begin{equation}
|\Psi _{0,L}^{(m)}\rangle =\mathcal{N}\sum_{\left\{ s\right\}
}q^{\sum_{j=1}^{L}js_{j}}|s_{1},s_{2},...,s_{L}\rangle ,  \label{gs}
\end{equation}%
where $m=\sum_{i}S_{i}^{z}$ denotes the magnetization sector, $s_{i}=\pm 1/2$%
, $\left\{|s_{1},s_{2},...,s_{L}\rangle\right\}$ is the basis where $S^{z}$ is diagonal,
and $\mathcal{N}$ is the normalization factor, which is found to be
\begin{equation}
\mathcal{N}=\left( \sum_{\left\{ s\right\} }q^{2\sum_{j=1}^{L}js_{j}}\right)
^{-1/2}.  \label{N}
\end{equation}%
The sums over spin configurations $\{s\}$ appearing in eqs.~$(\ref{gs})$ and $%
(\ref{N})$ are to be taken in the sector of magnetization $m$. The parameter
$q$ is defined by the following expressions
\begin{equation}
h_{c}=\frac{1}{4}(q-q^{-1})\,\,\,\,{\text{and}}\,\,\,\,\Delta =\frac{1}{2}%
(q+q^{-1}).  \label{q}
\end{equation}

Before describing our results we present the definition of concurrence, which is the 
measure of entanglement used throughout this paper. As shown in Refs.~\cite{Wootters:97,Wootters:98}, 
there is a one-to-one correspondence between concurrence and entanglement of formation~\cite{Bennet:96}, 
the latter being a strictly increasing function of the first. 
The concurrence $C_{12}$ for a pair of qubits labelled as 1 and 2 is 
defined by~\cite{Wootters:97,Wootters:98}
\begin{equation}
C_{12} = \text{Max}\left(\lambda_{1}-
\lambda_{2}-\lambda_{3}-\lambda_{4},0\right),
 \label{cdef}
\end{equation}
where the $\lambda_{i}$ are the square roots, in decreasing order, of the
eigenvalues of the operator 
\begin{equation}
R\equiv \rho_{12}(\sigma_{y}\otimes
\sigma_{y})\rho^{\ast}_{12}(\sigma_{y}\otimes\sigma_{y}).
\label{Rdef}
\end{equation}
In Eq.~(\ref{Rdef}), $\rho_{12}$ denotes the density matrix, which can be
either pure or mixed,  for the pair of qubits 1 and 2, and
$\rho^{\ast}_{12}$ its complex conjugate in the the standard basis
$\{|++\rangle,|+-\rangle,|-+\rangle,|--\rangle\}$. In a system with more than two
qubits, $\rho_{12}$ is obtained by tracing the density operator over the other qubits.
The concurrence ranges from 0, implying an unentangled state, to 1, in which the two qubits 
are maximally entangled.

As mentioned, the critical point given in Eq.~($\ref{hcrit}$) is invariant under changes in 
the size of the chain. Actually, as pointed out in Ref.~\cite{alcaraz}, two sites are enough 
to identify the critical field. From the point of view of entanglement, the phase transition 
can indeed be observed by looking at jumps in the concurrence for arbitrary pairs of spins. 
For $h<h_c$, the system is in the ferromagnetic phase, with vanishing concurrence for all 
the pairs. At the critical point, a wave function in the sector 
of magnetization $m=0$, which is entangled, becomes part of the ground state, but its 
mixture with the other sectors destroys entanglement. As $h$ crosses the critical value, 
the ground state becomes non-degenerate, with components only in the sector $m=0$, resulting in a 
jump in the concurrence for any pair of spins. This kind of non-analytical behavior in 
first-order transitions has been found  in Ref.~\cite{ibose}, where 
magnetization plateaus are associated to jumps in the entanglement. In our case, 
numerical computation of concurrence up to $24$ spins shows that the 
jumps are always observed in $h=h_c$ regardless the number $L$ of sites, in agreement 
with Ref.~\cite{alcaraz}. 


\begin{figure}[h!]
\centering
{\includegraphics[angle=0,scale=0.38]{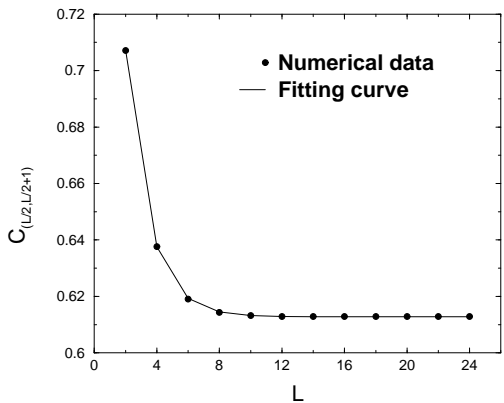}}
\caption{\label{f1} Concurrence for the central pair of spins as a function of the 
number $L$ of sites. The data have been obtained by taking the anisotropy $\Delta=1.36$ and  
the magnetic field $h=0.5$. As we can see, there is an exponential stabilization 
of the concurrence in the value $C \approx 0.61$, with the fitting function given by  
$C = A_0 + A_1\,{\textrm{exp}}(-A_2 L)$, where $A_0 = 0.612792$, $A_1 = 0.359898$ and $A_2 =
0.669476$.}  
\end{figure}
\begin{figure}[h!]
\centering
{\includegraphics[angle=0,scale=0.38]{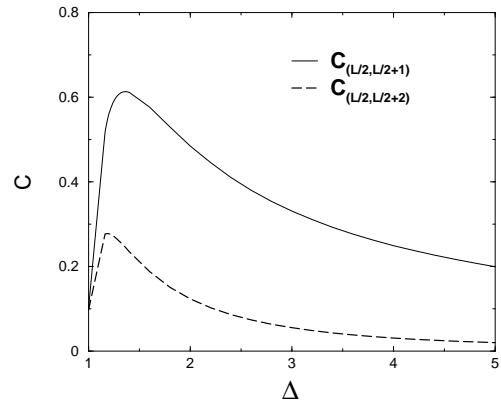}}
\caption{\label{f2} Concurrence for the pairs $(L/2,L/2+1)$ and $(L/2,L/2+2)$ as a function 
of the anisotropy $\Delta$. The data have been obtained by taking a chain with $22$ sites and  
the magnetic field $h=5.0$. The maximum value of $C_{(L/2,L/2+1)}$ occurs for $\Delta = 1.36$ and 
of $C_{(L/2,L/2+1)}$ for $\Delta= 1.18$.}  
\end{figure}

Concerning the distribution of entanglement along the chain, it is important to observe 
that the quantum domains break the ground state translation invariance, which implies 
that entanglement will depend not only on the distance of the spins, but also on their 
positions in the chain. The result is a concentration of pairwise entanglement in
the central region, with the spins in the center forming the most entangled pairs. 
In fact, for a system with even number $L$ of sites, it is possible 
to obtain a high degree of concurrence for the central pair, 
which evolves exponentially towards a stabilization value as we 
increase $L$. For instance, we show in Fig.~\ref{f1} that this concurrence stabilizes in 
$C\approx 0.61$ for $\Delta=1.36$ and $h=0.5$. This exponential behavior towards the 
stabilization is observed for all the pairs and the stabilization value is quickly  
obtained for large $\Delta$ and magnetic fields near $h_c$.
In fact, the stabilization value is strongly dependent on the anisotropy. In 
Fig.~\ref{f2} we show how the concurrences for the pairs $(L/2,L/2+1)$ and 
$(L/2,L/2+2)$ change as a function of $\Delta$ in a chain with $22$ sites. Note that, 
differently from the antiferromagnetic XXZ model without the quantum domains~\cite{Li:03}, 
the maximum of $C$ occurs for values of anisotropy greater than the isotropic value $1$.


On the other hand, the study of the concurrence as a function of the magnetic field suggests that 
the stabilization value is independent of the field for large 
chains. 
For example, we show in Table~\ref{t1} the concurrrence for the central pair 
as a function of $L$, for anisotropy $\Delta=1.5$ and several values of $h$. 
As it can be seen, there is  
clearly indication to an approximately constant value $C=0.5974378$ for any value of $h >h_c$ 
assuming that we are not extremely far from the critical field $h_c$. As we will show in Section~\ref{sce}, 
this numerical result leads to an analytical derivation of a general expression 
for the concurrence between arbitrary spins since, as given by Eq.~($\ref{gs})$, the exact ground 
state of the model is known at the critical magnetic field for a general $L$. 

\begin{table}[hbt]
\centering
\begin{tabular}{|c||c|c|c|}
\hline
$L$ & $h=0.6$         & $h=5.0$         & $h=20.0$        \\ \hline \hline
 6  & \,0.599509933\, & \,0.607452653\, & \,0.607652241\, \\ \hline
 8  & \,0.597825866\, & \,0.600881874\, & \,0.601276968\, \\ \hline
10  & \,0.597506824\, & \,0.598224627\, & \,0.598334941\, \\ \hline
12  & \,0.597449660\, & \,0.597593622\, & \,0.597617282\, \\ \hline
14  & \,0.597439774\, & \,0.597466481\, & \,0.597471030\, \\ \hline
16  & \,0.597438106\, & \,0.597442834\, & \,0.597443658\, \\ \hline
18  & \,0.597437830\, & \,0.597438641\, & \,0.597438785\, \\ \hline
20  & \,0.597437785\, & \,0.597437921\, & \,0.597437945\, \\ \hline
22  & \,0.597437778\, & \,0.597437800\, & \,0.597437804\, \\ \hline
24  & \,0.597437776\, & \,0.597437780\, & \,0.597437781\, \\ \hline
\end{tabular}
\caption[table1]{Concurrrence for the central pair $C_{L/2,L/2+1}$ as a function of the chain 
size $L$ for $\Delta=1.5$ and $h=0.6$, $5.0$, and $20.0$. Note that, 
as $L$ is increased, $C_{L/2,L/2+1}$ tends to the stabilization value $0.5974378$ independently of $h>h_c$.}
\label{t1}
\end{table}

\section{Concurrence from the ground state wave function with vanishing magnetization}

\label{sce}

In this section, we shall derive an analytical expression, holding in the kink-type phase, 
for the concurrences between arbitrary spins along the chain. As will be shown, this formula 
comes from the analysis of the entanglement present in one of the degenerates ground states at $h=h_c$, i.e.  
the wave function with magnetization $m=0$. Thus, let us suppose that the system is prepared in the 
vanishing magnetization state $|\Psi_{0,L}^{(0)}\rangle$, given by Eq.~$(\ref{gs})$. 
Then, the corresponding density operator is
\begin{equation}
\rho _{T}=\mathcal{N}^{2}\sum_{\{s\}\{s^{^{\prime
}}\}}q^{\sum_{j=1}^{L}js_{j}+\sum_{i=1}^{L}is_{i}^{^{\prime
}}}|s_{1}s_{2}...s_{L}\rangle \langle s_{1}^{^{\prime }}s{^{\prime }}%
_{2}...s{^{\prime }}_{L}|,  \label{rhot}
\end{equation}%
where the sums over $\{s\}$ and $\{s^{^{\prime }}\}$ are performed in the
sector $m=0$. The reduced density operator for a particular pair of spins in the sites $A$
and $B$ can be written as 
\begin{equation}
\rho =Tr_{\hat{A},\hat{B}}(\rho _{T}),  \label{rho}
\end{equation}%
with $Tr_{\hat{A},\hat{B}}$ denoting trace over all degrees of freedom
except the spins in the positions $A$ and $B$. Therefore the matrix elements
of $\rho $ are 
\begin{eqnarray}
\rho _{(s_{A},s_{B});(s_{A}^{^{\prime }},s_{B}^{^{\prime }})}=\mathcal{N}%
^{2}\,q^{A(s_{A}+s_{A}^{^{\prime }})}q^{B(s_{B}+s_{B}^{^{\prime
}})} \times \nonumber \\
\times \sum_{\{s\}} q^{2\left(\sum_{j=1}^{A-1}js_{j}+\sum_{j=A+1}^{B-1}js_{j}+\sum_{j=B+1}^{L}js_{j}
\right) },  \label{rhomat}
\end{eqnarray}%
where $s_{A}$, $s_{B}$, $s_{A}^{^{\prime }}$, and $s_{B}^{^{\prime }}$ represent 
the spin in the corresponding positions $A$ and $B$. The
sum over $\{s\}$ stands for the set of $(L-2)$ spin configurations $%
|s_{1}s_{2}...s_{(A-1)}s_{(A+1)}...s_{(B-1)}s_{(B+1)}...s_{L}\rangle $,
whose magnetization is such that 
\begin{eqnarray}
&&\sum_{j=1}^{A-1}s_{j}+\sum_{j=A+1}^{B-1}s_{j}+%
\sum_{j=B+1}^{L}s_{j}+s_{A}+s_{B}=0,  \label{condm01} \\
&&\sum_{j=1}^{A-1}s_{j}+\sum_{j=A+1}^{B-1}s_{j}+%
\sum_{j=B+1}^{L}s_{j}+s_{A}^{^{\prime }}+s_{B}^{^{\prime }}=0.
\label{condm0}
\end{eqnarray}%
The conditions (\ref{condm01}) and (\ref{condm0}) ensure that we are working in the sector of
magnetization $m=0$. Writing $\rho $ as a matrix, we have
\begin{equation}
\rho =\mathcal{N}^{2}
\left( \begin{array}{cccc}
q^{A+B}~\Omega _{1} & 0 & 0 & 0 \\
0 & q^{A-B}~\Omega _{0} & \Omega _{0} & 0 \\
0 & \Omega _{0} & q^{-A+B}~\Omega _{0} & 0 \\
0 & 0 & 0 & q^{-A-B}~\Omega _{-1}%
\end{array} \right)
\label{pmat}
\end{equation}
with the functions $\Omega _{i}$ $(i=-1,0,1)$ given by
\begin{equation}
\Omega _{i}=\sum_{\{s_{i}\}}q^{2\left(
\sum_{j=1}^{A-1}js_{j}+\sum_{j=A+1}^{B-1}js_{j}+\sum_{j=B+1}^{L}js_{j}%
\right) },  \label{Ss}
\end{equation}
where the sums over the $(L-2)$ spin configurations ${\{s_{i}\}}$ are
required to obey the restrictions
\begin{equation}
\sum_{\{s_{i}\}}\Longrightarrow \left(
\sum_{j=1}^{A-1}s_{j}+\sum_{j=A+1}^{B-1}s_{j}+\sum_{j=B+1}^{L}s_{j}\right)
+i=0.  \label{condSs}
\end{equation}
We are then able to compute the concurrence $C$, by
means of Eq.~$(\ref{cdef})$, for the spins in the sites $A$ and $B$, which gives
\begin{equation}
C=2\mathcal{N}^{2}\left( \Omega _{0}-\sqrt{\Omega _{1}~\Omega _{-1}}\right) .
\label{finalC}
\end{equation}%
Hence we have obtained a simple and very general expression to analyse concurrences in the model. 
In fact, as we discussed in Section~\ref{cc}, the ground state in the kink-type quantum phase is 
given by the wave function with $m=0$ and the corresponding concurrences are independent of the 
magnetic field for large chains (see Table~\ref{t1}). 
Therefore, concurrences obtained from the wave function with vanishing magnetization at $h=h_c$, given 
by Eq.~(\ref{finalC}), turn out to hold also in the kink-type phase, where $h>h_c$. In other words, entanglement  
in a large chain, computed from the $m=0$ ground state wave function, does not change as the magnetic field is 
increased. Eq.~(\ref{finalC}) displays a good property of being defined in terms of sums and not in terms of matrices, 
which are difficult to deal with when we have a large number of
sites. It is worth mentioning that this expression holds as a good approximation even in the case of 
small number $L$ of sites assuming that the magnetic field is close to the critical value $h_c$.
Agreement with numerical simulations is observed, providing support to the
exponential behavior, with respect to $L$, of the pairwise concurrences along the chain.

\section{The region $\Delta<1$}

\begin{figure}[h!]
\centering
{\includegraphics[angle=0,scale=0.45]{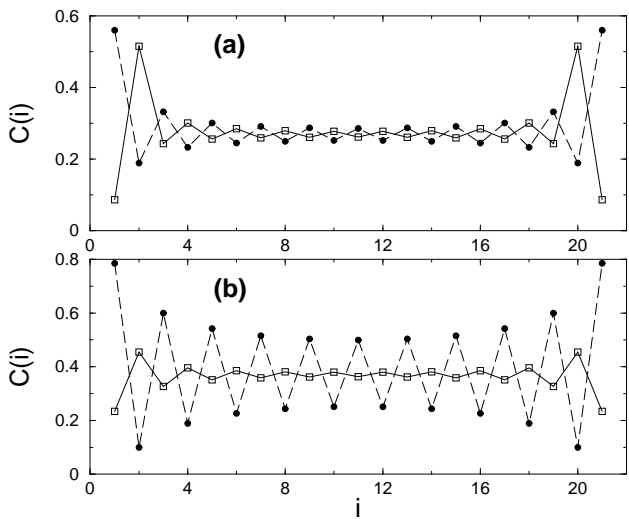}}
\caption{\label{f3} Distribution of entanglement for $\Delta<1$ in a chain of $L=22$ sites.
$C(i)$ represents the concurrence between the sites $i$ and $i+1$, where $i=1,...,L-1$.
The long-dashed lines with circles are associated to $h=0$ and the solid lines with squares
to $h=3.0$. In (a) we show the results for the gapless region for $\Delta=0.5$ and in (b) concurrence is
plotted in the gapful phase for $\Delta=-1.5$.}
\end{figure}

As discussed above, the concurrence for any pair of spins, for $\Delta\ge 1$,
presents an exponential behavior towards a constant value in the kink-type phase. Moreover, the
distribution of nearest neighbors concurrences along the chain exhibits a peak at the center, decaying
as we get nearer the boundary. However, as $\Delta$
is changed to less than the critical value $1$, entanglement properties become rather different.
As shown in Fig~\ref{f3}, the distribution of concurrences for nearest pairs are now characterized by
an oscillating behavior, with a sequence of peaks along the chain. Observe that the oscillations become
sharper as the anisotropy $\Delta$ is decreased and that the effect of the boundary field is a displacement of
the peaks, with a damping of the concurrences. 
The peaks can physically be explained by the tendency to
antiferromagnetic ordering as $\Delta<1$, which is observed from the negative sign of the 
two-point correlation functions  $<S_z(i)\, S_z(i+1)>$ for nearest neighbor spins $i$ and $i+1$.
Thus, for $h=0$ the spins $(1,2)$ tend to align antiferromagnetically, 
trying to form a bell state, and similarly the spins $(3,4)$, $(5,6)$, ..., $(L-1,L)$ (for $L$ even). 
The concurrences for these spins are greater than for the pairs $(2,3)$, $(4,5)$, ..., $(L-2,L-1)$ which,  
despite also exhibiting a tendency to antiferromagnetic alignment, are less correlated. When
the domain walls are considered, the boundary spins are governed by the quantum domains, favoring 
the correlations of the pairs $(2,3)$, $(4,5)$, ..., $(L-2,L-1)$, which leads to a dislocation 
of the peaks, in addition to a reduction of their sharpness. In the next section, we will show that 
the critical anisotropy $\Delta=1$ can also be very well characterized by looking at multiparticle 
entanglement properties of the model.

\section{Multipartite Entanglement}

Let us discuss now some multiparticle entanglement properties of the model, which will be seen to provide support 
to the results presented before. It is important to emphasize that, for multiparticle systems, we do not have 
a unique measure of entanglement. Then it turns out that a number of measures have been proposed~\cite{mw,Vedral:97,
Coffman:00,Vidal:00,Wong:01,Biham:02,Wei:03,Barnum:04}  
and different measures can be related to distinct physical aspects 
of multipartite entanglement. We shall consider here the so-called global entanglement, which is a scalable 
measure for pure states introduced by Meyer and Wallach in Ref.~\cite{mw}. For a general quantum state 
$|\psi\rangle$ in the Hilbert space $({\bf C}^2)^{\otimes L}$, the global entanglement is
\begin{eqnarray}
Q(|\psi\rangle) = \frac{4}{L} \sum_{j=1}^{L} D\left(|u^{(j)}\rangle, |v^{(j)}\rangle \right),
\label{mewa}
\end{eqnarray}
where the vectors $|u^{(j)}\rangle$ and $|v^{(j)}\rangle$, which are elements of the Hilbert space 
$({\bf C}^2)^{\otimes L-1}$, are defined by
\begin{eqnarray}
&&|u^{(j)}\rangle = t_j(-) |\psi\rangle = \sum_{x=1}^{2^{L-1}} u^{(j)}_x |x\rangle, \nonumber \\ 
&&|v^{(j)}\rangle = t_j(+) |\psi\rangle = \sum_{y=1}^{2^{L-1}} v^{(j)}_y |y\rangle.
\end{eqnarray}
with $t_j(b)$, for $b\in \{+,-\}$, denoting the following operation over basis vectors 
$|b_1 b_2 ... b_L\rangle$ of $({\bf C}^2)^{\otimes L}$:
\begin{eqnarray}
t_j(b)|b_1 b_2 ... b_L\rangle = \delta_{b b_j} |b_1 ... {\hat b_j} ... b_L\rangle,
\label{opmw}
\end{eqnarray}
In Eq.~(\ref{opmw}), the symbol $\,{\hat \frac{}{}}\,$ means absence. 
The function $D\left(|u^{(j)}\rangle, |v^{(j)}\rangle \right)$ 
in Eq.~(\ref{mewa}) is the norm-squared of the wedge product of $|u^{(j)}\rangle$ and 
$|v^{(j)}\rangle$
\begin{eqnarray}
D\left(|u^{(j)}\rangle, |v^{(j)}\rangle \right) = \sum_{x<y} |u^{(j)}_x v^{(j)}_y - u^{(j)}_y v^{(j)}_x |^2.
\label{dmw}
\end{eqnarray} 
It has recently been shown in Ref.~\cite{gb} that the global entanglement can be expressed in a simple 
way in terms of the one-qubit reduced density operator of the system
\begin{eqnarray}
Q(|\psi\rangle) = 2\left(1 - \frac{1}{L}\sum_{j=1}^{L} {\textrm{Tr}}(\rho_j^2) \right), 
\label{qbre}
\end{eqnarray}
where $\rho_j$ is the density matrix for the spin $j$ after tracing out the rest. 
From Eq.~(\ref{qbre}) we can numerically compute the global entanglement in our model. Since the 
measure is defined only for pure states, we shall always be referring to the kink-type phase, where 
$h>h_c$. As a first property, and in agreement on the results for the concurrence, $Q$ is seen to be 
independent of the magnetic field if we are not too far from the critical point. This motivates the 
analytical computation, as performed in the case of the concurrence in Section~\ref{sce}, of the 
global entanglement from the sector $m=0$ of the exact ground state 
Eq.~(\ref{gs}). The one-qubit reduced density matrix for a site $A$ in this case reads
\begin{equation}
\rho_A =\mathcal{N}^{2}
\left( \begin{array}{cc}
q^{A}~\Omega _{1/2} & 0 \\
0 & q^{-A}~\Omega_{-1/2}
\end{array} \right)
\label{predmw}
\end{equation}
where $\mathcal{N}$ is the normalization factor given by Eq.~(\ref{N}) 
and the functions $\Omega _{i}$ $(i=-1/2,1/2)$ are defined by
\begin{equation}
\Omega _{i}=\sum_{\{s_{i}\}}q^{2\left(
\sum_{j=1}^{A-1}js_{j}+\sum_{j=A+1}^{L}js_{j}\right) },  \label{Ssmw}
\end{equation}
with the sums over the $(L-1)$ spin configurations ${\{s_{i}\}}$ 
obeying the restrictions
\begin{equation}
\sum_{\{s_{i}\}}\Longrightarrow \left(
\sum_{j=1}^{A-1}s_{j}+\sum_{j=A+1}^{L}s_{j}\right)+i=0.  \label{condSsmw}
\end{equation}
Thus the global entanglement $Q$ can analytically be expressed by
\begin{eqnarray}
Q(|\psi\rangle) = 2\left(1 - \frac{\mathcal{N}^{4}}{L}\sum_{j=1}^{L} \left(q^{2j}\Omega_{1/2}^2 + 
q^{-2j}\Omega_{-1/2}^2 \right)\right)\hspace{-0.1cm}.
\end{eqnarray}
The expression above can be used to compute, in a very good approximation, the global entanglement 
for large chains in the kink-type phase assuming that we are not extremely far from the critical 
point, since a very high antiparallel boundary magnetic field can considerably increase $Q$. 
\begin{figure}[h!]
\centering
{\includegraphics[angle=0,scale=0.42]{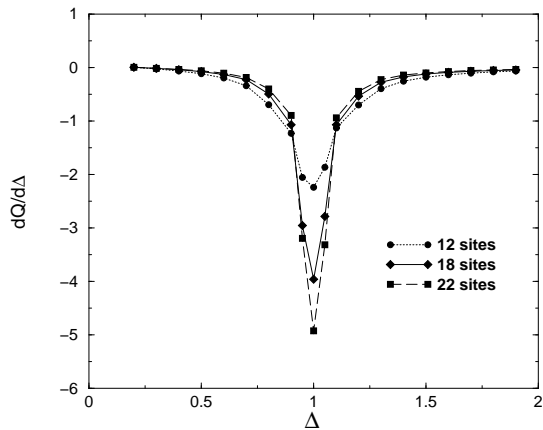}}
\caption{\label{f4} Derivative of the global entanglement $Q$ with respect to the anisotropy $\Delta$.
The magnetic field is set to $h=1.0$. Note the minimum in the critical value $\Delta=1$.}
\end{figure}

A further nice porperty of the global entanglement is that it presents an inflection point 
which characterizes the transition from the kink-type phase to the regime $\Delta<1$. In fact 
computing, from Eq.~(\ref{qbre}), the derivative of $Q$ with respect to the anisotropy, at a fixed number of sites,  
we observe the existence of a minimum exactly in $\Delta=1$, which gets more pronounced as 
we increase the number of sites. This behavior is 
explicitly displayed in Fig.~\ref{f4}. This characterization of the transition is 
rather attractive since, based on a multipartite measure of entanglement, it reflects a qualitative 
change in the variations of the ground state wave function of the system. Related characterizations 
of quantum phase transitions in terms of the global entanglement can be found in the 
Ref.~\cite{Somma:04} for the transverse field Ising model and in Ref.~\cite{Lambert:04} for 
the single-mode Dicke Hamiltonian.

\section{Conclusion}

In this paper, we have presented a discussion of how a nontrivial geometric interface, 
the ferromagnetic domain wall, can 
modify the pattern of entanglement in the spin-$1/2$ anisotropic Heisenberg chain. 
As shown, the quantum domains can be seen as a mechanism to generate pairwise entanglement  
in a model with Heisenberg ferromagnetic couplings, which is usually unentangled. Moreover we have seen that 
the amount of entanglement generated can be relatively large, specially for the central pair of spins. These 
effects are consequence of the first-order quantum phase transition induced by the presence of the 
domain walls. As a further nice result, we have been able to compute an analytical general
expression for the concurrence of large chains thanks to the numerical observation that,
in that case, entanglement is independent of the boundary magnetic field. Finally, we have detected different
patterns for the entanglement in the kink-type and $\Delta<1$ regimes. From the point of view of 
pairwise entanglement, these regions are identified by a different distribution of entanglement along the chain. 
Moreover these distinct phases have also been shown to be characterized by looking at the problem from the 
point of view of multiparticle entanglement. By considering 
the derivative of the global entanglement with respect to the anisotropy we showed that a minimum occurs 
exactly at the critical point $\Delta=1$. Applications of this kind of phase characterization in other models 
and the possible role of other multipartite entanglement measures are left as open problems for future investigation.

\section*{Acknowledgements}

This work was supported by the Brazilian agencies CNPq (F.C.A., A.S., M.S.S.) and FAPESP (F.C.A.). 
The research at the University of Waterloo was undertaken, in part, thanks to funding from the 
Canada Research Chairs Program (Michel Gingras) and at the University of Toronto thanks, in part, 
to the DARPA-QuIST program (Daniel A. Lidar). M.S.S. also thanks D.A. Lidar and S. Bandyopadhyay 
for enlightening discussions and for the reading of the present manuscript.

\end{document}